\documentclass[11pt, reqno,oneside]{amsart} 

\usepackage{amsmath}
\usepackage{amssymb}
\usepackage{amsfonts}
\usepackage{amsthm}
\usepackage{graphicx}
\usepackage{geometry}
\usepackage[utf8]{inputenc}
\usepackage{enumerate}
\usepackage{enumitem}
\usepackage[english]{babel}
\usepackage{bm}
\usepackage{pdfpages}
\usepackage{float} 
\usepackage{latexsym}
\usepackage{color}
\usepackage{hyperref}
\usepackage[numbers]{natbib}
\usepackage{caption}
\usepackage[foot]{amsaddr}
\usepackage{soul}
\usepackage[font=small,labelfont=bf,tableposition=top]{caption}
\usepackage[toc,page,titletoc]{appendix}

\DeclareCaptionLabelFormat{andtable}{#1~#2  \&  \tablename~\thetable}
\numberwithin{equation}{section}

\definecolor{questions}{rgb}{0.89, 0.35, 0.13}
\definecolor{todo}{rgb}{0.0, 0.45, 0.73}
\definecolor{comments}{rgb}{0.47, 0.53, 0.6}

\newcommand{\E}[1]{\mathbb{E}\left[#1\right]}
\newcommand{\Prob}[1]{\mathbb{P}\left(#1\right)}

\newcommand{\specialcell}[2][c]{%
	\begin{tabular}[#1]{@{}c@{}}#2\end{tabular}}

\newtheorem{example}{Example}

 \allowdisplaybreaks
 
\begin{document}

\title[]{
Modelling preventive measures and their effect on generation times in emerging epidemics
}

   
\author[]{ Martina Favero $^1$}
\address{$^1$ Department of Mathematics, Stockholm University}
\author[]{ Gianpaolo Scalia Tomba $^2$}
\address{$^2$ Department of Mathematics, University of Rome Tor Vergata}
\author[]{ Tom Britton $^1$}


\newpage

\maketitle

\begin{abstract}
We present a stochastic epidemic model to study the effect of various  preventive measures, such as uniform  reduction of contacts and transmission, vaccination, isolation, screening and contact tracing, on a disease outbreak in a homogeneously mixing community.
The model is based on an infectivity process, which we define through stochastic contact and infectiousness processes, so that each individual has an independent infectivity profile.
In particular, we monitor variations of the reproduction number and of the distribution of generation times. 
We show that some interventions, i.e. uniform reduction and vaccination,  affect the former while leaving the latter unchanged, whereas other interventions, i.e. isolation, screening and contact tracing, affect both quantities. 
We provide a theoretical  analysis of the variation of these quantities,  and we show that, in practice,  the variation of the generation time distribution can be significant and that it can cause biases in the estimation of reproduction numbers.  
The framework, because of its general nature, captures the properties of many infectious diseases, but particular emphasis is on COVID-19, for which numerical results are provided.

\end{abstract}

\keywords

\section{Introduction}

While the reproduction number is usually seen as a dynamic quantity, changing over the course of an epidemic, the generation time distribution is often seen as a static object. 
For example, during the recent COVID-19 pandemic, 
major efforts have been made to continuously estimate reproduction numbers, while often outdated estimates of the generation time distribution have been  employed in the estimation.
The distribution of the generation time, which we define here as the time between the infection of a  secondary case and the infection of the corresponding primary case, 
is not an intrinsic property of an infectious disease, on the contrary, 
it depends on the environment and on the behaviour of the individuals among whom the disease spreads. Consequently, the generation time distribution can be  subject to variations, 
for example, 
recent studies hint that generation times of SARS-CoV-2 have significantly shortened during the course of the pandemic \cite{ali2020,hart2021,scarabel2021,torneri2021}.
When doing inference, it is important to investigate the extent of these variations and take them into account, when substantial.

The significance of the generation time distribution  stems from its relation to the initial exponential growth rate of the epidemic and the basic reproduction number, or equivalently the current growth/decline rate and the current reproduction number.
In fact, by observing the incidence, the growth rate is often estimated and used together with the generation time distribution to derive the reproduction number by means of the Euler-Lotka equation, see section \ref{sect:gen_time} for more details. 
Therefore, using a current estimate of the growth rate  together with a  generation time distribution estimated before preventive measures were in place could lead to a biased estimate of the  reproduction number. 
Estimating the reproduction number $R$ in an ongoing epidemic is crucial for efficient control of the epidemic.
If for example $R= 1.25$, the overall number of contacts of infectious
individuals must be reduced by at least $20\%$ to stop the epidemic from increasing ($(1-0.2)R = 1$). Similarly, a community currently having a high level of restrictions and current reproduction number $R = 0.8$ may relax restrictions as long as the number of contacts
by infectious individuals does not increase by more than $25\%$ ($ (1+0.25)R = 1)$. Biases in
estimates of $R$ may hence lead to improper conclusions regarding control measures.

Several issues arise in connection with estimation of the generation time distribution. 
To begin with, in practice  it is usually not possible to observe all of the secondary cases caused by a cohort of primary cases, therefore biases due to missing cases arise. 
Furthermore, moments of infection are rarely observed and surrogate time points are used, such as time of symptom onset.
When the epidemic is exponentially growing, short generation times are overrepresented among the observed ones.  
The generation time may vary during the course of the epidemic because of significant changes in the number of susceptible individuals, for example near the peaks of incidence.
When performing inference it is important to take the above mentioned issues  into account, see e.g.  \cite{champredon2015,svensson2007,scaliatomba2010,svensson2015,britton2019}.
Another problem, of a different nature, is that the generation time may vary because of changes in the behaviour of individuals, in particular  due to preventive measures introduced to reduce the spread of the disease. 

The latter problem has received some attention, see 
for example \cite{ali2020,scarabel2021, torneri2021} in connection with the COVID-19 pandemic,
but, unlike the other mentioned problems, has not been extensively studied yet.
It is not uncommon that studies, which focus on the important task of estimating other quantities, use an outdated estimate of the generation time distribution which does not represent the current situation,   because of  changes in the level of preventive measures among different time periods or among different locations. 
For example, many influential and highly cited studies, such as \cite{ ferretti2020, flaxman2020},  use the estimate of the generation time distribution of SARS-CoV-2 from early 2020 in Shenzhen, China \cite{bi2020}. 
See also \cite{hart2021}, where this problem is highlighted,  and references therein.
Other influential studies, e.g.  
\cite{kissler2020},  use instead estimates of the generation time distribution \cite{wallinga2004} of SARS-CoV-1 which despite being similar to SARS-CoV-2, resulted in milder preventive measures. 

The choice of using outdated estimates of the generation time distribution to obtain much-needed estimates of other quantities, such as reproduction numbers, is  understandable, however, it may lead to biases.
That is why in this paper we focus on
illustrating, both theoretically and in a COVID-19 example,  how various preventive measures change the generation time distribution in emerging epidemics and we draw attention to some scenarios in which it would be preferable to replace the initial estimate of the generation time distribution with an up-to-date estimate that takes into account the non-negligible effect of interventions.
A related problem is the effect of asymptomatic individuals on generation times, which we briefly discuss in Subsection \ref{sect:asympt_bias}.

We focus on the initial phase of an epidemic, that is when depletion of susceptible individuals is negligible.
As the epidemic progresses, some of the potentially infectious contacts of infected individuals will be with already infected or recovered (assumed immune) individuals and thus the effective reproduction number will decrease. The reduction is proportional to the fraction of individuals that have already been infected and hence only slowly changing in the initial phase of the epidemic. 
Determining the length of this phase depends on the specific outbreak, and thus goes beyond the scope of this paper, but can be done for example by observing that the infection rate in the beginning of the epidemic is close to constant up until  a certain percentage of individuals have been infected, say  $5\%$. The time it takes to reach this threshold identifies the initial phase. 
Our analysis of the generation time distribution can also be easily transferred to other phases of the epidemic in which the fraction of susceptible individuals does not change significantly. Continuing the previous example,  also later in the epidemic the infection rate is  close to constant over a period where at most $5\%$ get infected.
This period is usually shorter around the peak,
for example it might take a few months for the first $5\%$ to get infected  and later in the epidemic, around the peak, a few weeks for  $5\%$ more to get infected. 
Note that considering later phases of the epidemics would allow one to take into account  naturally-acquired immunity, but not waning of immunity if waning is observed over a longer period than the periods mentioned above.

We present, in Section \ref{sect:model},  a general stochastic epidemic model for the spread of infectious diseases
with a structure that is particularly indicated to  analyse the impact of preventive measures on the distribution of generation times and  reproduction numbers. 
The key feature of the model that facilitates this analysis is a random infectivity profile with a random time at which the contact activity of an infectious individual is reduced. We define this random time to be equal to the time of symptoms onset, when no preventive measures are in place, and equal to the time of detection (that can occur by various means, including symptoms onset), when interventions are in place. We assume that for each individual the rate of symptoms onset is proportional to infectiousness, see Subsection \ref{sect:random_sympt} for a  discussion on this assumption and further details.

In particular,  we consider the following interventions: homogeneous reduction of contacts and transmission, vaccination, isolation, screening and contact tracing. 
Clearly all these preventive measures aim at reducing the reproduction number, however, not all of them have an effect on generation times.  
The analysis in this paper shows that homogeneously reducing infectivity, by reducing contacts or transmission, as well as vaccination do not change the generation time distribution, see Subsections \ref{sect:uniform_reduction} and \ref{sect:vaccine}.
On the contrary, reducing the contact activity of, or isolating, symptomatic individuals does have a significant impact on generation times, see Section \ref{sect:isolating_symptomatic}. For example, in the COVID-19 scenario illustrated in Section \ref{sect:illustration}, the natural intuition that this preventive measure shortens generation times is confirmed.
Furthermore, our analysis shows that other
interventions affecting the generation time distribution are those  aiming at expediting the time at which an infectious individual is isolated, or reduces their contact activity, 
such as screening and contact tracing.
In particular, in the COVID-19 scenario with isolation of detected individuals, if the population is screened uniformly, that is, each individual is randomly tested at a certain constant rate,  
generation times are shortened, and  
if contact tracing is put in place, the shortening becomes substantial. 
Using a pragmatic approach,  aimed at providing a general analysis of the impact of contact tracing at the population level,  we make an approximation of the contact tracing mechanism and 
we provide a rigorous derivation of the rate at which  individuals are contact traced. See Subsection \ref{sect:shortening_detection} and the supplementary material for a detailed discussion on screening and contact tracing. 

In Section \ref{sect:all_interventions} we summarise the results in a general formula for the generation time distribution  that takes into account the cumulative effect of all interventions.
The model, the analysis and the general formula for the generation time distribution (sections \ref{sect:model} - \ref{sect:interventions}) are valid under general conditions and thus can be used to study various infectious diseases by defining a suitable infectivity profile and parameters.
In Section \ref{sect:illustration}, we 
tune the model to fit a COVID-19 scenario 
to illustrate  the theoretical general results in a  realistic example, and, perhaps more importantly, 
to  investigate the extent of the impact of interventions on the generation time distribution and reproduction number in a framework that allows for further extensions.

 \vspace{0.3cm}
\begin{table}[h]
\scalebox{0.9}{
\begin{tabular}{|l|l|}
\hline 
Infectivity process  and infectivity function                 & $\lambda=\{\lambda(t)\}_{t\geq 0}, \beta$  \\
Infectiousness process            & $X=\{X(t)\}_{t\geq 0}$         \\
Contact process                      & $C=\{C(t)\}_{t\geq 0}$        \\
Contact rate before and after symptom onset / detection & $C_1,C_2$ \\
Time of contact activity reduction and its conditional rate      & $\tau,\alpha_\tau$     \\
Time of symptoms  onset, its conditional rate and corresponding parameter & $T_S,\alpha_S,a_S $ \\
Probability of asymptomatic infection & $p^{asy}$ \\
Time of detection   and its conditional rate   & $T_D,\alpha_D$     \\
Time of screening and its rate & $T_{scre},\alpha_{scre},\sigma$ \\
Time of contact tracing and its conditional rate & $T_{CT},\alpha_{CT}$  \\
Probability of successful contact tracing & $p$ \\
Other quantities related to contact tracing & $\alpha_{CT1},f, \alpha_{CT2}, a_{CT2},d$ \\
Reduction fractions of contacts and transmission & $\rho_C,\rho_X$\\
Reduction fractions due to vaccination   & $\rho_V $\\
 \quad and to isolation of symptomatic and detected individuals & $\rho_S,\rho_D$ \\
Fraction of vaccinated individuals & $v$ \\
Relative susceptibility (vaccination response) & $A, a$ \\
Relative infectivity (vaccination response) & $B, b$ \\
\hline
\end{tabular}
}
\vspace{0.3cm}
\caption{Notation summary}
\label{table:notation}
\end{table}

\section{The stochastic model}
\label{sect:model}
The model we present can be interpreted  as a special case of the very general epidemic models in e.g. \cite{ball1995,svensson2007}, tailored to consider preventive measures and to study their impact. 
A summary of the notation, including stochastic processes, random variables, parameters and functions is given in Table \ref{table:notation}. 
We adopt the classical assumption that the population is homogeneously mixing.  This is a natural assumption, despite it being a simplification of reality, as it allows an analysis which is valid
for the numerous wide-spread models that are based on the very same assumption.  
Future investigation could consist of removing this assumption and repeating the analysis  for more complex models based on networks.
In this model, we also assume that  all individuals are  equally susceptible at the start of the epidemic, whereafter they might be infected and later removed. 
On the contrary, infectious individuals are not equally infectious, they have infectivity profiles which are independent and identically distributed, as in \cite{ball1995,svensson2007}. That is, the infectivity profile of an infectious individual is the realization of a stochastic process $\lambda$, the infectivity process,
which is the product of a contact process and an infectiousness process, described in the following.
This allows modelling some inhomogeneity in the population, in fact, although identically  distributed,  individuals' contact behaviours and infectiousness can vary. 
It is known \cite{ball1995} that, when the initial phase of an epidemic is considered, and thus depletion of susceptible individuals is negligible,
 this epidemic model corresponds to a Crump-Mode-Jagers  branching process \cite{jagers1975}. 
 This allows using results from the theory of branching processes to analyse the epidemic model.

\subsection{Infectiousness process}
The infectiousness profile of an infectious individual is an independent realization of the 
 continuous-time, $[0,1]$-valued stochastic process $X=\{X(t)\}_{t\geq0} $, 
the infectiousness process. 
Given  $X$, if a contact between the infectious individual with infectivity profile $X$ and a susceptible individual occurs at time $t$ (since the infection of the infectious individual), then infection of the susceptible individual happens with probability $X(t)$. 
In the following, we mention two possible choices for the infectiousness process.

\begin{example}[Infectiousness profiles with a deterministic shape]
\label{example:deterministic_shape}
Let $X_1$ be a random variable with values in $[0,1]$, $X_2$ a random variable with values in $\mathbb{R}_{> 0}$, and let  $h:\mathbb{R}_{\geq 0}\to [0,1]$ be a deterministic function.
Then a possible construction of the infectiousness process is $X(t)=X_1 h( t X_2^{-1})$. This is based on the assumption that the  infectiousness profile has the same shape for all individuals, given by the function $h$. Furthermore, since not all individuals are equally infectious, each individual is associated to a pair of random variables distributed as $(X_1,X_2)$, which determine the strength and the location of the peak of infectiousness. 
\end{example}

\begin{example}[Infectiousness profiles in the SIR model]
\label{example:SEIR}
In the well known SIR stochastic model, the infectiousness process corresponds to $X(t)=x \mathbb{I}_{[0,I]}(t)$, where $x$ is the probability of infection  and 
 $I$ is the (random) length of the infectious period.
\end{example}

\subsection{Contact process}
The contact activity of an infectious individual is a realization of the continuous-time, $\mathbb{R}_{\geq 0 }$-valued, stochastic process
$C=\{C(t)\}_{t\geq0}$,  
the contact process. An individual's  contact activity  may depend on their infectiousness profile, but is assumed to be independent of other individuals' contact activities. 
Given $C$, 
an infectious individual with contact activity $C$ has contacts with other individuals at the time points of an inhomogeneous Poisson process with intensity $C(t)$,  time being measured since the infection of the infectious individual. 

We assume that the contact process has the following form  
\begin{equation}
\label{def:contact_process}
    C(t)=\begin{cases}
    C_1, \text{ if }  t\leq \tau\\
    C_2 , \text{ if }  t>\tau
    \end{cases}
    =
    C_1 \mathbb{I}_{t\leq \tau}
    +C_2 \mathbb{I}_{t> \tau},
\end{equation}
where $\tau$ is a random time at which the initial base contact activity $C_1$ is reduced  to $C_2$.
The  $\mathbb{R}_{\geq 0}$-valued random variables $C_1$ and $C_2$, with $C_2\leq C_1$, are possibly dependent on each other, but are assumed to be independent of anything else.
We assume that 
the random time  $\tau$  has conditional rate  $\alpha_\tau(t)$, at time $t$, 
given the contact activities, $C_1,C_2$, and 
given the infectiousness processes until time $t$, that is, given
$\mathcal{X}_t$
which corresponds to the natural filtration associated to the infectiousness process.  
This means that the conditional distribution of $\tau$ is of the form 
    \begin{equation}
    \label{eq:tau_distr}
    G_\tau(t)=
    \Prob{\tau>t\mid \mathcal{X}_t, C_1,C_2}=
    \exp\left(-\int_0^t\alpha_\tau(u) du \right)
    .
    \end{equation}

Having defined a contact process of the form \eqref{def:contact_process} allows enough flexibility to describe realistic scenarios we are interested in analysing, while enabling a solid mathematical formulation. 
In fact, as shown in the next sections, 
different definitions of $\alpha_\tau, C_1,C_2$, 
allow modelling contacts in several scenarios, with and without preventive measures. 
In particular, $\tau$ can be defined to model for example the time of symptoms onset or, more generally, the time of detection by symptoms onset or by other preventive measures, e.g. screening and contact tracing. More details on $\tau$ are provided  in Section  \ref{sect:time_sympt} and Subsection \ref{sect:shortening_detection}, where  modelling of symptoms onset and  detection are respectively addressed.

\subsection{Infectivity process}
Infectivity profiles of infectious individuals are independent realizations of the continuous-time, $\mathbb{R}_{\geq 0 }$-valued stochastic process  $\lambda=\{\lambda(t)\}_{t\geq0}$, with  
    $$
    \lambda(t)=C(t)X(t).
    $$
The infectivity process $\lambda$ combines the effects of the infectiouness process and the contact process. In a  susceptible population, 
given $\lambda$, an infectious individual with infectivity profile $\lambda$  infects other individuals at the time points of an inhomogeneous Poisson process with intensity $\lambda(t)$, time being measured since the infection of the infectious individual.
This is equivalent, by properties of Poisson processes, to the description in the previous subsections: the infectious individual has contacts at the time points of an inhomogeneous Poisson process with intensity $C(t)$, and a contact at time $t$ results in an infection with probability $X(t)$. Note that 
the process that counts the infectious contacts,
i.e. the counting process $N=\{N(t)\}_{t\geq 0}$, with random intensity $\lambda$, is a Cox process, also known as doubly-stochastic Poisson process, see e.g. \cite{snyder1991}.
We have described the role of the infectivity process in a  completely susceptible population, instead, if some individuals are not susceptible, contacts with those cannot result in infection. However, in the initial phase of an epidemic, which is what this paper focuses on, depletion of susceptible individuals is negligible and thus the above description of the infectivity process is valid.

\section{Some quantities of interest}

\subsection{Reproduction numbers}
In this framework, the infectivity function, i.e. 
the average rate at which a typical infected individual infects others  in a fully susceptible population,  is simply the expectation of the infectivity process, that is 
\begin{align}
\label{eq:beta}
\beta(t)=\E{\lambda(t)}&=
\E{C_1 X(t) G_{\tau}(t)
+
C_2 X(t) 
\left(1-G_{\tau}(t)\right)},
\end{align}
where  $G_{\tau}(t)$ is  defined in \eqref{eq:tau_distr}.
Therefore, the average number of individuals infected by a typical infectious individual, 
the reproduction number, is 
\begin{equation}
\label{eq:R0}
R=\int^\infty \beta(t) dt 
=R^{(1)} +R^{(2)} ,
\end{equation}
where 
    $$
    R^{(1)}=
\E{\int_0^\infty
C_1 X(t) G_{\tau}(t)
dt}
    $$
is the average number of individuals infected by a typical infectious individual before reducing their base contact activity (e.g. before symptoms onset, or before detection),
and
    $$
    R^{(2)}=
    \E{ \int_0^\infty
C_2 X(t) 
\left(1-G_{\tau}(t)\right)
dt}$$
is the average number of individuals infected by a typical infectious individual after changing their contact activity.

The above formulas for $R,R^{(1)}$ and $R^{(2)}$  give a theoretical means of comparison of the effect of various preventive measures, 
in particular,  considering $R^{(1)}$ and $R^{(2)}$ separately can be useful to better understand strengths and limits of a certain intervention. For example, by simply isolating infectious individuals after they show symptoms, $R^{(2)} $ can be significantly reduced, however, if no other preventive measures are in place, $R^{(1)}$ remains unchanged and thus  $R$ cannot be reduced below $ R^{(1)} $. 
However, often in practice, the reproduction number cannot be estimated directly by formulas of the type above and estimates rely instead on another approach, which is described in the following subsection.

\subsection{Generation time distribution}
\label{sect:gen_time}
A  valuable  tool to estimate the reproduction number is  the Euler-Lotka equation  which
relates the reproduction number $R$ to the, usually more easily observable, Malthusian parameter, the exponential growth rate $r$, see e.g. \cite{wallinga2007}. 
As shown in \cite{britton2019}, in the initial phase of an epidemic, the incidence quickly approaches exponential growth, with rate $r$,  
and thus the Euler-Lotka equation, arising from population dynamics, applies
    \begin{equation*}
    1= R \int_0^\infty e^{-r t} g(t) dt .
    \end{equation*}
This equation links the reproduction number and the growth rate
through the function $g$, the generation time distribution,  which, in the framework of this paper, as in \cite{svensson2007}, is expressed as
    \begin{equation}
    \label{eq:gen_time}
    g(t)=\frac{\E{\lambda(t)}}{\E{\int_0^\infty\lambda(u)du }} =
    \frac{\beta(t)}{R} .
    \end{equation}
It is because of its appearance in the Euler-Lotka equation that the generation time distribution has a crucial role in inference. In fact, knowing $g$ allows deriving $R$ from $r$ or vice versa. 
As pointed out in \cite{svensson2007}, the intuitive definition of the generation time as the time between a primary and a secondary case has several mathematical counterparts, and it is thus important to recall that $g$ in the Euler-Lotka equation corresponds to the following definition.
Consider a cohort of infectious individuals (referred to as primary cases) and all of their secondary cases. Choose one of the secondary cases uniformly at random and measure the time between the infection of the secondary case and the infection of the corresponding primary case. The probability density function of this random time is equal to $g$ in \eqref{eq:gen_time}, see \cite{svensson2007} for more details. Note that this sampling procedure is size-biased, that is, a primary case associated to more secondary cases is more likely to be chosen compared to one associated with fewer secondary cases. 
For this reason,  choosing first a primary case uniformly at random and then one of its secondary cases (also uniformly at random)  would yield a different generation time distribution, $\E{\frac{\lambda(t)}{\int_0^\infty\lambda(u)du}}$,  which is not of interest here and should not be confused with  \eqref{eq:gen_time}.

This paper focuses on the study of variations of the generation time distribution due to preventive measures. 
Let us explain why these variations are worth  being studied. 
Imagine an epidemic evolving without preventive measures, with basic reproduction number $R_0$, generation time distribution $g_0$ and growth rate $r_0$. 
When interventions are introduced, $R_0, g_0, r_0$ are replaced by $R_E, g_E, r_E$.
Each of the triplets above satisfies the Euler-Lotka equation. When doing inference  before interventions, $r_0$ is observed and $R_0$ is derived using  $g_0$.  The same should be done when  various interventions are put in place, that is, $r_E$ and $g_E$ give $R_E$.
However, 
if we observed $r_E$ and used $g_0$ instead of $g_E$,  the resulting reproduction number would be  biased.
The aim of this paper is to study how the generation time distribution varies, from $g_0$ to $g_E$, when preventive measures are introduced, and how that affects the estimate of the reproduction number.

\section{Modelling symptoms and the neutral scenario}
\label{sect:time_sympt}

The stochastic model described in Section \ref{sect:model}  can 
capture several scenarios,
including the neutral scenario in which no preventive measures are in place. In this case,  the random time $\tau$, at which the contact activity of an individual is reduced,  is assumed to be equal to the time of symptoms onset $T_S$.
In fact, a natural reduction of the contact activity can occur at symptoms onset, even without preventive measures, for example, 
even if  isolation of symptomatic people is not required, having flu-like symptoms naturally reduces contact activity.

Furthermore, by defining $\tau=T_S$ we can study  all  those interventions that do not affect the time at which the contact reduction occurs, see Subsections \ref{sect:uniform_reduction}, \ref{sect:vaccine} and \ref{sect:isolating_symptomatic}. Whereas, for those  interventions that aim at expediting the time at which the contact reduction occurs, e.g. by implementing other means of detection besides symptoms onset, $\tau$ needs to be defined differently, i.e. as the time of detection, see Subsection \ref{sect:shortening_detection}.

This section is dedicated to $\tau=T_S$, 
in particular, Subsection \ref{sect:random_sympt} is dedicated to mathematical definition of $T_S$, the underlying biological assumptions,  and the resulting formulas, which will be crucial in the following sections, while Subsection \ref{sect:asympt_bias} contains a comment on the effect of asymptomatic cases.

\subsection{Random time of symptoms onset}
\label{sect:random_sympt}
Let $T_S$ be the time between infection and symptoms onset of an infectious individual.
Consider the model of Section \ref{sect:model} with $\tau=T_S$.
We assume that $T_S$ has conditional rate, given $\mathcal{X}_t$,
\begin{equation}
\label{eq:rate_sympt}
\alpha_S(t)=a_S  X(t),
\end{equation}
where $a_S\in \mathbb{R}_{>0}$.
In this way the distribution of the time to symptoms onset of an individual depends on their infectiousness process and is independent  of  their contact process. Furthermore, the above assumption implies that, at any given time, the higher the infectiousness, the higher the rate of symptoms onset. Therefore, it is more likely for a symptomatic infectious individual to show symptoms in proximity to the peak of their infectiousness,
 in line with what is observed for COVID-19, see e.g. \cite{he2020,hart2021b}, and with the viral load of patients with pandemic H1N1 2009 virus infection peaking on the day of onset of symptoms \cite{to2010}.
 Furthermore,  
note that in this model an infectious individual is not necessarily symptomatic, even if highly infectious.
We call  an infectious individual who never shows symptoms asymptomatic (to not be confused with an individual who has not shown symptoms yet, called pre-symptomatic),  this corresponds to  $T_S=\infty$. 
The probability that an individual with infectiousness process $X$ is asymptomatic is
    $$
    \Prob{T_S=\infty\mid X}
    =
    e^{-\int_0^\infty a_S X(u) du}.
    $$
This means that the higher the total infectiousness, $ \int_0^\infty  X(u) du$, the higher the probability of the individual being symptomatic.
More generally, the probability that a typical individual is asymptomatic is                     $$p^{asy}=\Prob{T_S=\infty}
    =\E{e^{-\int_0^\infty a_S  X(u) du}}.$$
Making different biological assumptions on symptoms onset would correspond to a different mathematical definition of $T_S$.  For example, $T_S$ could be defined as the time at which the infectiousness process exceeds a certain threshold. 
From now on \eqref{eq:rate_sympt} is assumed. This is not only a realistic assumption, but also mathematically convenient,   yielding explicit formulas  
  for $R^{(1)}$ and $R^{(2)}$,  the average number of individuals infected by a typical infectious individual before and after symptoms onset, respectively.
See the supplementary material for more details.

\subsection{Biases due to asymptomatic individuals}
\label{sect:asympt_bias}
Usually the generation time distribution is estimated from data related to symptomatic cases. That is, instead of the real generation time distribution $g$, the generation time distribution of symptomatic cases, $g^{sy}$, is estimated,  and in turn used to estimate the  reproduction number, leading to biases, see e.g. \cite{park2020, torneri2021}. In this subsection we briefly analyse how these two distributions are related.

Let 
$\beta^{asy}(t)=\E{\lambda(t)\mid T_S= \infty}$ and 
$\beta^{sy}(t)=\E{\lambda(t)\mid T_S< \infty}$
be the average infectivity rates of asymptomatic and symptomatic individuals respectively.
Then the average infectivity rate is
\begin{equation*}
    \beta(t)=p^{asy} \beta^{asy}(t) + (1-p^{asy})\beta^{sy}(t).
\end{equation*}
Let $R^{asy}=\int_0^\infty\beta^{asy}(t) dt$ and 
$R^{sy}=\int_0^\infty\beta^{sy}(t) dt$
be the average numbers of individuals infected by a typical asymptomatic infectious individual and by a typical symptomatic individual respectively. 
It is straightforward to calculate that 
    $$
    R=p^{asy}R^{asy}+ (1-p^{asy})R^{sy}.
    $$
The formula above confirms the obvious intuition that 
the 
higher the fraction $p^{asy}$ of asymptomatic individuals, the higher the impact of asymptomatic individuals and thus, 
when data on asymptomatic individuals are unavailable,
estimates of $R$ might be biased, unless a correction is made.
Furthermore, $g^{sy}(t)=\frac{1}{R^{sy}}\beta^{sy}(t)$ and the generation time distribution of asymptomatic individuals is 
$g^{asy}(t)=\frac{1}{R^{asy}}\beta^{asy}(t)$, thus
    $$
    g(t)= 
    q^{asy}
    g^{asy}(t) 
    +(1-q^{asy}) 
    g^{sy}(t)
    $$
where  
    $$
    q^{asy}=
    \frac{p^{asy}R^{asy}}{p^{asy}R^{asy}
    + (1-p^{asy})R^{sy}} 
    $$ 
indicates how much asymptomatic transmission affects the overall generation time distribution.

\section{Interventions}
\label{sect:interventions}

By varying the contact process $C$, the infectiousness process $X$, or the random time $\tau$, several types of intervention, and their effect on generation times and reproduction number, can be studied. 
In particular, the interventions analysed in this paper are  grouped in the following categories:

\begin{itemize}
    \item
    Homogeneous reduction of contact level (e.g. physical distancing, lockdown) 
    \item Homogeneous reduction of  transmission probability  (e.g. face masks)
    \item
    Vaccination
    \item Isolation of symptomatic individuals
    \item Screening
    \item Contact tracing
\end{itemize}
Note that the latter two interventions aim at expediting the time at which an individual is detected and at lowering their contact activity after detection, while the remaining ones focus on lowering the infectivity process, without affecting the detection time. 
In the following subsections we analyse the different types of preventive measures, considering one  at a time and comparing it with the no-interventions neutral scenario. 
In Subsection \ref{sect:all_interventions}, the cumulative effect of all interventions is considered and  a general formula is provided.

\subsection{Homogeneous reduction of contact activity  or transmission probability}
\label{sect:uniform_reduction}
This subsection is dedicated to the analysis of two types of preventive measures that lead to the reduction of the infectivity process by a multiplicative factor.

The first class of preventive measures that we consider consists of those measures, such as physical distancing or lockdown, that reduce the contact process by a factor $\rho_C$, which is a $[0,1]$-valued random variable.
This means that each individual reduces their contact activity by a factor which is an independent realization of $\rho_C$.

The second class instead consists of those measures, such as introducing face masks, that reduce the infectiousness process by a factor $\rho_X$, which is a $[0,1]$-valued random variable. This means that the infectiousness profile of each individual is reduced by a factor which is an independent realization of $\rho_X$.

These measures are homogeneous in the population, that is, we are assuming that each individual is recommended/required to follow the same measures independently of their situation. The random variables $\rho_C$ and $\rho_X$ are thus independent of the infectivity process.
While the measures are homogeneous, the individual response is variable, and thus we use random variables, instead of deterministic constants, to represent the variability in the personal adherence.  
 
Both classes of measures, albeit in different ways, have the same effect on the infectivity process, which is  reduced by a multiplicative factor, i.e.
    $
    \lambda_E(t) = \rho_C\rho_X  \lambda(t),
    $
which yields 
    $
    \beta_E(t) =\E{\rho_C\rho_X} \beta(t)$ and 
    $ R_E=\E{\rho_C\rho_X} R_0.
    $
Therefore,  the preventive measures considered in this subsection affect the reproduction number, while leaving the generation time distribution unchanged. 
 
\subsection{Vaccination}
\label{sect:vaccine}
In  this subsection we complement the underlying epidemic model with a  vaccination model  as in e.g. \cite{becker1998,ball2007}.  
We assume that a fraction $v$ of the population receives a vaccine, before the epidemic starts. While this is a simplification of reality, it allows an analysis of the generation time distribution in a population that is partly vaccinated and a comparison with the scenario without interventions.
Assume that each vaccinated individual has a random response to the vaccine, determining the reduction in susceptibility and (if infected) infectivity. The response is described by the $[0,1]$-valued random variables $A$ and $B$,  the relative susceptibility and  the relative infectivity  respectively. 
This means that a vaccinated individual with response $A,B$ has a probability of getting infected reduced by $A$ compared to the probability of getting infected without vaccine, and if ever infected, their infectivity is reduced by a factor $B$. 
Note that we are assuming that the vaccine may reduce the infectiousness profile by a multiplicative factor $B$, without changing its shape, see the supplementary material for a discussion on this assumption. 

As in the previous subsection, the infectivity function, as well as  the reproduction number, is reduced by a multiplicative factor, $\E{\rho_V}$, and therefore  the generation time distribution is unchanged. 
 For more details see the supplementary material.

Immunity, or partial immunity, from disease exposure has the same type of effect as vaccination, that is, it  reduces the reproduction number without changing the generation time distribution. In fact,
as explained in the Introduction, in this paper we are considering a period of time in which the fraction of immune individuals does not change significantly, thus the reasoning around the effect of immune individuals is identical to the one around vaccinated individuals. 

Finally, we remark that the effect of immunity waning in vaccinated individuals is not considered here. This goes beyond the scope of the model, since we only focus on dynamics over  shorter time periods, i.e. the phases mentioned in the Introduction,  rather than on the long term  evolution which  usually needs to be considered  when analysing the effect of waning.

\subsection{Isolating symptomatic individuals}
 \label{sect:isolating_symptomatic} 

As mentioned in Section \ref{sect:time_sympt}, in a scenario without preventive measures, the initial contact activity $C_1$  of an individual is naturally reduced to $C_2$ at the time of symptoms onset, $T_S$. 
If symptomatic individuals are recommended or required to isolate, then the contact activity after symptoms onset, $C_2$, is further reduced by a factor $\rho_S\in [0,1]$, ideally close to $0$. 
This preventive measure can be easily included in the model and its effect on the generation time distribution and the reproduction number can be studied by analysing \eqref{eq:beta} and \eqref{eq:R0}. 

The basic reproduction number, in the no-interventions scenario is 
 $R_0=R_0^{(1)}+R_0^{(2)}$. Now assume that symptomatic individuals are required/recommended to isolate, while no other preventive measures are in place. 
 It is straightforward to see that the reproduction number becomes  $R_E= R_0^{(1)}+\rho_S R_0^{(2)}$.
Therefore, by simply isolating symptomatic individuals,  the reproduction can be lowered to a minimum of $R_0^{(1)}$, which corresponds to a scenario in which symptomatic individuals are completely isolated and do not have any contact with others. In fact, isolating symptomatic individuals has no impact on the amount of pre-symptomatic or asymptomatic transmission, which defines $R_0^{(1)}$.
The   generation time distribution, by \eqref{eq:beta} and \eqref{eq:R0}, becomes
    $$
    g_E(t)=
    \frac{1}{R_0^{(1)}+\rho_S R_0^{(2)}}
    \E{C_1 X(t) G_{T_S}(t)
    +
    \rho_S C_2 X(t) 
    \left(1-G_{T_S}(t)\right)}.
    $$
Therefore, isolating symptomatic individuals not only affects the reproduction number, but also, 
unlike the previously analysed interventions, 
changes the generation time distribution. 

If other detection measures are in place, the preventive measure discussed in this subsection could be improved by isolating detected individuals in addition to symptomatic individuals. See the next subsections for more details.

\subsection{Screening and contact tracing}
\label{sect:shortening_detection}

In the previous subsections, the time at which an individual reduces their contact activity, $\tau$, was assumed to coincide with the time of symptoms onset. 
While isolating symptomatic individuals or generally reducing the infectivity process homogeneously in the whole population does not affect $\tau$, 
other preventive measures instead aim at reducing $\tau$.
  
Interventions such as screening and contact tracing act by expediting the time at which an infectious individual is discovered to be infectious.
This scenario can be modelled by letting $\tau=T_D$, the time of detection, and
    $$T_D=\min\{T_S,T_{scre},T_{CT}\}$$
where $T_{scre}$ is the time of  screening, $T_{CT}$ is the time of detection by contact tracing. 
In this framework, we assume that, given the infectiousness and the contact process up to time $t$, i.e. given $\mathcal{X}_t$ and $\mathcal{C}_t$, the conditional rate of detection at time $t$ since infection is 
    \begin{align}
    \label{eq:det_rate}
      \alpha_D(t)=\alpha_S(t)+\alpha_{scre}(t)+\alpha_{CT}(t)
    \end{align}
where $\alpha_S$, the conditional rate of symptoms onset, is defined in \eqref{eq:rate_sympt}, $\alpha_{scre}$, the screening rate, is defined in \eqref{eq:scre_rate} below, and $\alpha_{CT}$, the conditional rate of detection by contact tracing, is defined in \eqref{eq:CTrate} below.

We assume screening is performed as follows.
Random test are carried out so that the entire population, including infectious and non infectious individuals, is screened uniformly. That is, independently of their infectiousness and contact process, each individual is tested at a constant rate $\sigma$. At time $t$, given $X(t)$, the rate of detection by screening is thus 
    \begin{align}
    \label{eq:scre_rate}
    \alpha_{scre}(t)=\sigma \mathbb{I}_{X(t)>0},
    \end{align}
as an individual can only be detected when infectious.
This means that, if we consider a small time period of length $\epsilon$, any individual, infectious or not,  has approximately  a probability $\epsilon \sigma $ of being tested during that period, thus roughly a fraction $\sigma$ of the entire population is tested each day, if time is measured in days. 
Uniform screening is not often used in practice, and thus not representative of most real-world scenarios. 
The next section illustrates that the effect of uniform screening is moderate, unless 
$\sigma$ is large, which in practice requires a substantial screening effort. This motivates why uniform screening is not often implemented as a preventive measure.
In order to obtain a bigger impact, it is more efficient to direct testing towards individuals that are more likely to be infectious, as in contact tracing programs, rather than spreading it uniformly over the entire population. Being a more efficient strategy, this is also more representative of the real-world practice, for this reason we now include contact tracing in the model. 

Modelling contact tracing is notoriously challenging, see \cite{muller2021} for an extensive overview. 
Various modelling approaches are possible, see for example \cite{browne2015,browne2022}  for compartmental models, \cite{scarabel2021} for deterministic integral equations, \cite{muller2000} for an individual-based stochastic model; and important observations on data have been made, see e.g. \cite{sun2021}.
Nevertheless, simpler models that are wide-spread in practical applications rely on strong simplifying assumptions, while more accurate complex models are often intractable in practice. 
A complicating factor is that, because of contact tracing, infectious individuals are not independent of each other, 
for example, the infectivity profile of an infector might be truncated because one of their infectees develops symptoms quickly and triggers contact tracing leading to detection of the infector.
In order to exactly model contact tracing, it is necessary to keep track of the status of each single individual and of relations between individuals by building additional mathematical structure in the model. An example is given by \cite{muller2000}, where each individual in a stochastic SIRS model is associated to an id-number and with the id-number of their infector, this leads to complicated calculations and large simulations, even if the underlying epidemic model is rather simple.
The heterogeneity of individuals, which is modelled in this paper by the stochastic infectivity process, 
leads to further difficulties.
Therefore,  instead of using a similar approach, aiming at exact expressions and keeping track of each single individual, we focus on approximating the effect of contact tracing to provide insight at population level.

The first approximation that we make is to assume that the infectivity profiles of different individuals are  independent of each other, despite contact tracing. 
Each individual can be contact traced either through one of their infectees, with conditional rate $\alpha_{CT1}$, or through their infector, with conditional rate $\alpha_{CT2}$.
As a result of the approximation and of the assumptions described below, 
the rate at which an infectious individual is detected through contact tracing at a certain time $t$ since their infection, 
given $C_1$ and the infectiousness profile up to time $t$, i.e. given $\mathcal{X}_t$, 
 is
    \begin{align}
    \label{eq:CTrate}
    \alpha_{CT}(t)=
   \alpha_{CT1}(t)+\alpha_{CT2}(t)
    =
    p\int_0^{t-d} C_1 X(u) f(t-d-u)du + p a_{CT2} \mathbb{I}_{X(t)>0} ,
    \end{align}
where $p$ is the probability that contact tracing occurs successfully, $p a_{CT2}\mathbb{I}_{X(t)>0}$ is the rate, approximated by a constant, at which an infectious individual is detected through their infector, $f$  is the probability density function of the time between the moments an infectee is infected and detected, and $d$ is the (deterministic) contact tracing  delay.
 It is implied that, if $t\leq d$, the integral in \ref{eq:CTrate} is equal to zero and thus $\alpha_{CT1}(t)=0$.
 We assume that contact tracing stops after one step, which is a reasonable approximation of reality. In fact, it is unlikely that an individual is contact traced through their infectee who in turn has been contact traced, because this would usually take longer than the infectious period. This latter assumption allows deriving an explicit expression for the function $f$, which would not be possible otherwise,  see the supplementary material. Formula \ref{eq:CTrate} can be easily generalised to include a random contact tracing delay, as shown in the supplementary material.  
 Finally, assuming that the infector is detected at a constant rate is also an approximation of reality. In fact, this rate varies with time and depends on how long has passed since the infector was infected, while  remaining independent of the infectivity profile of the infectious individual under consideration. 
 A  non-approximated derivation of this rate is challenging, as explained in the supplementary material, and would require additional structure to be added to the model, which goes beyond the scope of this paper and is left for future work.

\subsection{The cumulative impact of all interventions}
\label{sect:all_interventions}
Finally, to summarise the analysis of this section,
we consider altogether the preventive measures that have been analysed so far and provide a general formula for the generation time distribution under the effect of all preventive measures. 
  When all preventive measures are in place, 
  and in particular, not only symptomatic individuals, but also detected individuals reduce their contact activity by a fraction $\rho_D$, 
the infectivity function becomes
    \begin{align}
    \label{eq:general_formula_beta} 
    &\beta_E(t)=
    \E{\rho_V\rho_C\rho_X} 
    \E{C_1 X(t) G_{T_D}(t)
    +
    \rho_D C_2 X(t) 
    \left(1-G_{T_D}(t)\right)},
    \\
&\text{with} \nonumber
    \\
    \label{eq:general_formula_G} 
    &G_{T_D}(t)=
     \exp\left(
     -a_S \int_0^t  X(u) du - (\sigma+ p a_{CT2}) (t \wedge I ) 
     - p C_1 \int_0^{t-d}   X(u)[1-y (t-d-u) ]du   
     \right), \\
     \label{eq:general_formula_y} 
     &y(t)= \E{\exp\left(-\int_0^t a_{S}X(u)du-\sigma (t \wedge I ) \right)}
     ,
    \end{align}
 where $\wedge$ indicates the minimum and $I$ is the length of the infectious period, i.e. $I=\int_{0}^{\infty}\mathbb{I}_{X(u)>0}du$.   
From the expressions above, it is clear that, while all interventions affect the reproduction number, 
$ R_E=\int \beta_E(u)du$,     
only some interventions are found to have an impact on
the generation time distribution, $g_E(t)=\frac{\beta_E(t)}{R_E}$,
 that is, isolation, screening and contact tracing.
 In Section \ref{sect:illustration} the expression above is used to illustrate  variations of the generation time distribution in a realistic example.
 
To relate our results to a model  used in numerous applications, we conclude this section with two examples concerning the well-known SIR model, which can be seen as a special case of our  general model. 

\begin{example}[Basic SIR model]
	\label{example:basicSIR}
Continuing with the SIR model of Example \ref{example:SEIR}, we recall that the infectivity process is of the form $\lambda(t)=  c \xi \mathbb{I}_{[0,I]} $, where $\xi$ is the probability of infection, $c$ is the contact rate and $I$ is the length of the infectious period, which is exponentially distributed with parameter $a_I$. 
In this framework, it is straightforward to show that 
$g(t)=a_I \exp(-a_I t)$, that is, 
 the generation time is exponentially distributed with parameter $a_I$, see e.g. \cite{svensson2015}.

\end{example}

\begin{example}[SIR model with interventions]
\label{example:SIRinterventions}
Consider a SIR model where the infectious period can be cut short by symptoms onset. That is, the length of the infectious period is 
$I'=\min\{I,T_S\}$. 
The time of symptoms onset has rate $\alpha_S(t)=a_S \xi \mathbb{I}_{[0,I]}$, thus
$G_{T_S}(t)=
\exp\left(-  a_S \xi (t\wedge I) \right)
$. 
It is then easy to show that the generation time is exponentially distributed with parameter $ a_S \xi+a_I$.
Note that in Example \ref{example:basicSIR} neither $c$ nor $\xi$  influence the generation time distribution. On the contrary, in this example, a higher $\xi$ lowers the mean generation time  by increasing the rate of symptoms onset, $a_S \xi$. 

Screening and contact tracing can be also considered in the SIR model and a formula 
for the generation time distribution can be easily obtained from formulas 
\ref{eq:general_formula_beta},
\ref{eq:general_formula_G}, and
\ref{eq:general_formula_y}, by simply plugging in $X(t)=\xi\mathbb{I}_{[0,I]}$, using that $I$ is exponential  and analytically computing the integrals. Straightforward but lengthy calculations show that the  generation time distribution in this example is a generalisation of a truncated (positive) Gumbel distribution, see the supplementary material for more details.

\end{example}

\section{Illustration: COVID-19 outbreak}
\label{sect:illustration}
In this section we tune the model to resemble a COVID-19 outbreak and illustrate the impact of preventive measures in such a framework.
While we use available evidence to define the parameters of the model and thus provide a realistic illustration of Covid-19 scenarios,
we do not directly use data in our study.
The results presented here are not  to be considered as proper estimates of generation times and reproduction numbers for the COVID-19 pandemic, but rather as a means to illustrate the extent of variation caused by interventions.

\begin{figure}[h]
	\centering
	\includegraphics[width=0.7\textwidth]{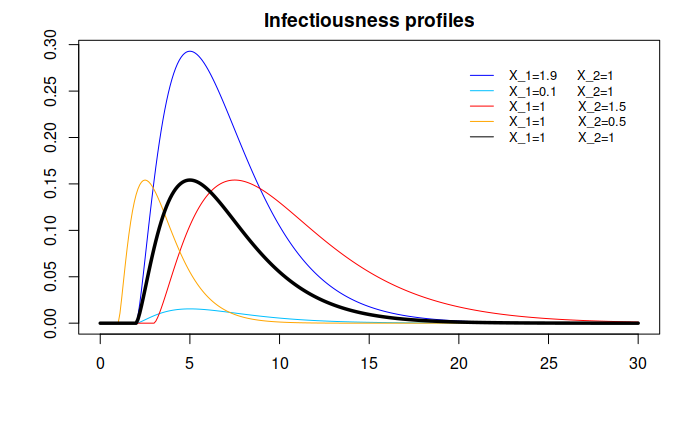}
	\vspace{-1cm}
	\caption{In black, the deterministic curve that shapes the infectiousness process. In colors, various infectiousness profiles, corresponding to the most extreme cases: in blue and light blue, the highest and lowest peak of infectiousness,  respectively; in red and orange the latest and earliest peaks of infectiousness,  respectively. }
	\label{fig:profiles}
\end{figure}

The infectiousness process is assumed to be of the form described in Example \ref{example:deterministic_shape} with the function $h(t)$ being a Gamma density,  with shape $2.5$ and  rate $0.5$, shifted by $2$, in line with  the analysis in \cite{he2020,hu2021}, and 
$X_1$ and $X_2$ are uniformly distributed in $[0.1,1.9]$ and $[0.5,1.5]$, respectively. 
We recall that each infectious individual has an infectiousness profile which corresponds to a realisation of the random infectivity profile, thus, in this case, to a realisation of the variables $X_1$ and $X_2$. 
In Figure \ref{fig:profiles}, some infectiousness profiles, corresponding to different realisations of $X_1$ and $X_2$ are plotted. In particular are plotted the underlying profile, corresponding to $X_1=1,X_2=1$, and the four profiles corresponding to the extreme cases $X_1=0.1,1.9, X_2=0.5,1.5$. 

In order to model superspreaders, the contact rate $C_1$ is assumed to have a Pareto II distribution with shape $2.1$, scale $5$ and minimum $0$. 
The contact rate after symptoms onset or detection is assumed to be $C_2=\rho C_1$.
For example, $\rho=1$ corresponds to no reduction of contact activity after symptoms onset  and  $\rho=0$ corresponds to complete isolation.
Furthermore $a_S=2$, as with this choice the fraction of asymptomatic individuals is around $\frac{1}{4}$, in line with \cite{alene2021}.
Monte Carlo integration is used to compute the expectations in \eqref{eq:general_formula_beta}, \eqref{eq:general_formula_y}.

\subsection{Reducing the contact activity of symptomatic individuals}
In this section we analyse the effect of reducing the contact activity of   symptomatic individuals by 
letting $\rho$ vary between $0$ and $1$ and reporting variations of the generation time distribution and reproduction numbers in  Figure \ref{SY} and Table \ref{SY}. 
We recall that the values presented here are not estimates, they are calculated using the formulas of Subsection \ref{sect:all_interventions} with the choice of parameters described at the beginning of this section.

When no interventions are in place, it is still unrealistic to expect no reduction in the contact rate due to symptoms, i.e. $\rho=1$, first row of Table \ref{SY} and blue curve in Figure \ref{SY}. In fact, an  individual with COVID-19 may  naturally reduce contact activity when developing symptoms, even if not explicitly recommended or required  to do so. This is simply because symptoms, especially if severe, are often associated with a tendency to reduce social activities, as for example is the case with seasonal flu for which no strict isolation measures are in place. 
Therefore, the scenario without interventions should correspond to one of the first rows of Table \ref{SY}, for example to $\rho=0.8$, as will be assumed in Subsection \ref{sect:bias}. Whereas, the ideal scenario in which all symptomatic individuals are completely isolated at symptoms onset corresponds to $\rho=0$.
It is evident that, as expected, isolating symptomatic individuals lowers the mean generation time. We conclude that the variation of the generation time due to reducing the contact activity of  symptomatic individuals may be, as in this example,  quite significant.  While the variation of the reproduction number is also significant, this preventive measure cannot bring the reproduction number below a certain threshold, 
even when  applied perfectly, i.e. $\rho=0$, because of pre-symptomatic and asymptomatic transmission.

\vspace{0.5cm}
\begin{minipage}[c]{0.6\textwidth}
	\centering
	\includegraphics[width=1\textwidth, height=0.7\textwidth]{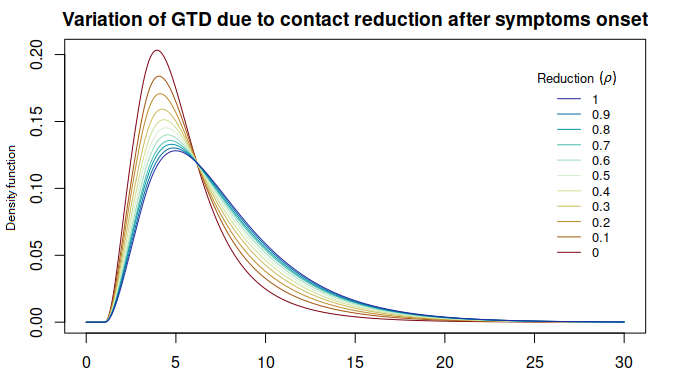}
\end{minipage}
\begin{minipage}[c]{0.4\textwidth}
	\scalebox{0.9}{
		\begin{tabular}{|p{6mm}|p{8mm}|p{8mm}|p{8mm}|p{8mm}|}
			\hline
			$\rho$
			&  $R$ 
			& $R^{(1)}$ 
			& $R^{(2)}$ 
			&  MGT \\
			\hline
			1&  4.54 & 1.73 & 2.81& 7.57\\
			0.9& 4.26 & 1.73& 2.53&  7.48\\
			0.8& 3.98 & 1.73& 2.25& 7.38\\
			0.7 & 3.70 & 1.73 & 1.97 & 7.28 \\
			0.6& 3.42 & 1.73& 1.69  & 7.15\\
			0.5& 3.14 & 1.73 & 1.41 & 6.99\\
			0.4& 2.86& 1.73& 1.13& 6.82\\
			0.3& 2.57&  1.73 & 0.84& 6.59\\
			0.2& 2.29& 1.73 & 0.56 &  6.31\\
			0.1& 2.01 & 1.73 & 0.28 &  5.96\\
			0 &1.73 & 1.73& 0&  5.48\\
			\hline
		\end{tabular}
	}
\end{minipage}
\captionlistentry[table]{}
\captionsetup{labelformat=andtable}
\captionof{figure}{ On the left hand side, the generation time distribution (GTD) for different values of $\rho$, the reducing fraction of contact activity at symptoms onset. No other interventions are in place, i.e. $\sigma=0,p=0$. The blue line  corresponds to no reduction ($\rho=1$) and the red line corresponds to complete isolation  ($\rho=0$) at symptoms onset. On the right hand side, the corresponding reproduction numbers (total, $R$, before symptoms, $R^{(1)}$, and after symptoms, $R^{(1)}$) and mean generation times (MGT). }
\label{SY}
\vspace{0.5cm}

\subsection{ Isolating symptomatic and screened infectious individuals }

Starting from the assumptions of the previous subsection with $\rho=0$, we now include screening. That means that when an individual is detected, through  symptoms or screening,  their contact activity ceases. 

In Figure \ref{SY_SCRE} and Table \ref{SY_SCRE} 
we report variations of the generation time distribution and of reproduction numbers due to the screening rate $\sigma$ varying between $0$ and $0.1$.
The impact of uniform screening is moderate,  mean generation times and reproduction numbers are not affected  as much as in the previous subsection.  Furthermore,  a rate  $\sigma=0.01,$ entails a high  effort in practice, as it  requires roughly $1\%$ of the population being screened each day. Higher values of $\sigma$ are hardly reachable in practice, and as we mentioned above, contact tracing or other targeted testing procedures might be more efficient than increasing the value of $\sigma$.

\subsection{ Isolating symptomatic and contact traced infectious individuals }
\label{sect:ill_CT}
In this subsection, we exclude screening, i.e. $\sigma=0$  and we consider contact tracing.
At symptoms onset or when they are contact traced, individuals are completely isolated,  i.e. $\rho=0$. 
The rate at which an individual is detected by  through-infector contact tracing is chosen to be $a_{CT2}=0.1$, the effect of this rate, is analogous to the effect of the screening rate and can thus be seen in the previous subsection, here instead we focus  on illustrating the effect of contact tracing through-infectees.  To this aim, 
we let the probability of successful contact tracing, $p$, vary between $0$ and $1$, which respectively correspond to the scenario in which no contacts an individual has made can be traced and the scenario in which all contacts can be traced. The results are reported in Figure \ref{SY_CT} and Table \ref{SY_CT}. 
The contact tracing delay is chosen to  be  $24$ hours, i.e. $d=1$. This is the time between the detection of an infectee and  the detection of their infector through contact tracing.

\vspace{0.5cm}
\begin{minipage}[c]{0.6\textwidth}
\centering
 \includegraphics[width=1\textwidth, height=0.7\textwidth]{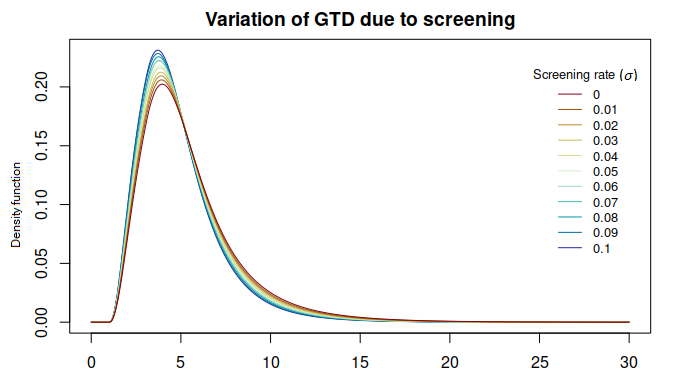}
\end{minipage}
\begin{minipage}[c]{0.4\textwidth}
\centering
\scalebox{0.9}{
\begin{tabular}{|p{8mm}|p{8mm}|p{8mm}|}
\hline
 $\sigma$ 
&  $R$   
&  MGT  \\
\hline
  0 & 1.73&   5.48 \\
 0.01 & 1.64 &  5.40 \\
  0.02& 1.57 & 5.31 \\
  0.03 & 1.47 & 5.25 \\
  0.04 & 1.39 & 5.20 \\
  0.05 & 1.32 &  5.13 \\
  0.06 & 1.26 & 5.07 \\
  0.07 & 1.20& 5.00 \\
  0.08& 1.14& 4.94\\ 
   0.09 & 1.08 & 4.90\\
  0.1 & 1.04 & 4.84 \\
\hline
\end{tabular}
}
\end{minipage}
\captionlistentry[table]{}
\captionsetup{labelformat=andtable}
\captionof{figure}{On the left hand side, the generation time distribution (GTD) for different values of $\sigma$, the screening rate. Detected individuals are completely isolated, i.e. $\rho=0$, and there is no contact tracing,  i.e. $p=0$. On the right hand side, the corresponding reproduction numbers  and mean generation times (MGT).}
\label{SY_SCRE}

\vspace{0.5cm}
\begin{minipage}[c]{0.6\textwidth}
\centering
\includegraphics[width=1\textwidth, height=0.7\textwidth]{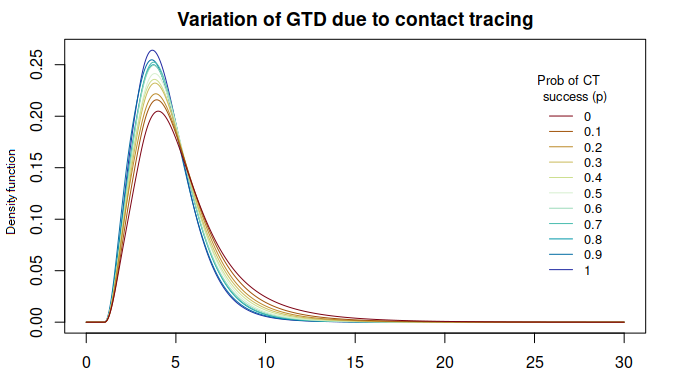}
\end{minipage}
\begin{minipage}[c]{0.4\textwidth}
\centering
\scalebox{0.9}{
\begin{tabular}{|p{8mm}|p{8mm}|p{8mm}|}
\hline
 $p$  
&  $R$  
&  MGT  \\
\hline
0 & 1.73   &    5.48\\
0.1 & 1.50 &  5.17 \\
0.2 & 1.45 &  4.98 \\
0.3 & 1.31 &  4.86 \\
0.4 & 1.21 &  4.75 \\
0.5 & 1.16 &  4.68 \\
0.6 & 1.10 &  4.60 \\
0.7 & 1.05 &  4.56 \\
0.8 & 0.94 &  4.52 \\
0.9 & 0.90 & 4.41 \\
1 & 0.87  &  4.38 \\
\hline
\end{tabular}
}
\end{minipage}
\captionlistentry[table]{}
\captionsetup{labelformat=andtable}
\captionof{figure}{On the left hand side,  the generation time distribution (GTD) for different values of $p$, the probability of successful contact tracing. Detected individuals are completely isolated, i.e. $\rho=0$, and there is no screening,  i.e. $\sigma=0$. On the right hand side, the corresponding reproduction numbers  and mean generation times (MGT).}
\label{SY_CT}

\newpage
 
\subsection{Biased estimates of reproduction numbers: an example}
\label{sect:bias}
In the previous subsections we have illustrated that the generation time distribution may vary significantly  due to  interventions when symptomatic and detected individuals (through screening or contact tracing) are isolated. In this subsection we illustrate how this variation may lead to significantly biased estimates of the reproduction number by presenting several scenarios, one without interventions, one with all interventions and others with some interventions.
See Table \ref{tab:scenarios} for a summary of the chosen parameters for the various scenarios and the corresponding variations. 

\vspace{0.3cm}
\begin{table}[h]
\centering
\begin{tabular}{|c|c|c|c|c|c|c|c|c|c|c|} 
\hline
 Isolation & Screening & \specialcell{Contact\\Tracing} & $\rho$ & $\sigma$ & $p$ & $r$   & MGT & $R$  & $\hat{R}$  \\ 
\hline
No & No & No  & 0.8       & 0      & 0 & 0.23   & 7.38 & 3.98 &  3.98 \\
Yes & Yes & Yes   & 0.2   & 0.01   & 0.7  & 0.09 & 6.17 &  1.67 & 1.84 \\
Yes & No & No   &   0.2   &  0  &  0 & 0.15 & 6.31 & 2.29 & 2.60 \\
Yes & Yes & No   &  0.2   &  0.01  & 0 & 0.14 & 6.29 & 2.23 & 2.53 \\
Yes & No & Yes   &   0.2   & 0 & 0.7 & 0.10 & 6.12 & 1.73 & 1.92  \\
\hline
\end{tabular}
\vspace{0.3cm}
\captionsetup{labelformat=simple}
\caption{For each  scenario the table reports  the  chosen  parameters (reducing fraction of contact activity $\rho$,  screening rate $\sigma$ and  fraction of known contacts $p$) and the resulting growth/decline rate $r$, mean generation time $MGT$ and reproduction numbers, $R$ which is estimated from the actual generation time distribution, and $\hat{R}$ estimated from the generation time distribution $g_0$ of the neutral scenario. }
\label{tab:scenarios}
\end{table}

\subsubsection*{Neutral scenario: no interventions}
Using the same underlying variables as in the previous subsections, we assume that in this neutral scenario without interventions, i.e. $\sigma=0,p=0$, symptomatic individuals reduce their contact activity by $20\%$, i.e. $\rho=0.8$. 
Imagine that at the beginning of an outbreak, when no preventive measure is in place, a growth rate, $r_0=0.23$, and a  generation time distribution, $g_0$ with mean $7.38$, are estimated. By the Euler-Lotka equation then an estimate of the basic reproduction number, $R_0=3.98$, can be obtained. 

\subsubsection*{All interventions in place}
Imagine that, during the same outbreak described in  the last paragraph, at a later time, preventive measures are introduced and another growth rate, $r_E=0.09$, is estimated. 
We assume that, in this scenario, 
detected individuals reduce their contact activity by $80 \%$, i.e. $\rho=0.2$, that roughly $ 1\%$ of the population is uniformly screened every day, i.e. $\sigma=0.01 $, and that contact tracing is performed in such a way that  $70\%$ of the total contacts are reported, i.e. $p=0.7 $. 

If we were to use the wide-spread assumption that the generation time distribution does not change during the course of the outbreak, we would use the
previously estimated generation time distribution, $g_0$, and the Euler-Lotka equation 
to obtain an estimate of $1.84$ for the new reproduction number  $R_E$. However, because of interventions, the  generation time distribution has in fact changed from  $g_0$ to $g_E$,  with a mean of $6.17$, which is calculated through the formulas in Subsection \ref{sect:all_interventions}.
Using the actual  generation time distribution yields the more precise value of $1.67$ for $R_E$. 
\\

Therefore, the variation of the generation time distribution between the neutral scenario and the scenario with interventions causes a non-negligible bias in the estimate of the reproduction number. 

The converse problem can also arise if estimates of the generation time distribution obtained under  stricter intervention regimes are used to provide estimates of the reproduction number when interventions are less strict or absent.  
As mentioned in the introduction, often studies use estimates of the generation time distribution which are based on data collected in China, where extensive contact tracing, including even enforced digital contact tracing, and severe isolation measures were in place. Using this estimate of the generation time distribution to estimate the reproduction number in other countries, where milder preventive measures are in place,
may lead to biases, analogously to the situation described above. Likewise, the estimate of the generation time distribution made  in one country at a certain time when stricter measures are in place, may lead to biases if used in that same country once those measures are eased. For example, it is relevant to take this issue into account when studying the effect of relaxing preventive measures.

The variation of the generation time distribution, and the corresponding bias in the reproduction number estimate, is not always significant and thus can be in some cases neglected, see for example Table \ref{tab:scenarios}, where various scenarios with various combinations of interventions are reported. 
In particular, the closer is the growth/decline rate to zero, the less the estimate of the reproduction number is sensitive to changes of the generation time distribution.

 \section{Concluding remarks}

 The contribution of this paper is threefold. 
 A general model, suitable to describe the spread of various infectious diseases under the effect of various preventive measures, is presented. Variations of reproduction numbers and generation times and related biases are analysed. General theoretical formulas are provided and applied in an illustration of a realistic COVID-19 scenario. 
 
 The generality of the model relies on the random infectiousness profile and on the random contact activity of  infectious individuals, which allows taking inhomogeneity into account. In this way,   the infectivity can vary not only over time, but also among different individuals, depending for example on the severity of the infection.
 In Section \ref{sect:illustration} we choose a distribution for the  infectiousness profile in line with the results in \cite{he2020,hu2021}, however the formulas in Subsection \ref{sect:all_interventions} can be used in other settings   for any random  infectiousness profile, which should be chosen to match appropriate characteristics. 
 Similarly, the random contact activity, which is reduced at a certain random time,  allows taking into account different social behaviours of infectious individuals, including for example super spreaders. 
 The model allows a variety of possible choices of explicit forms for distributions, dependencies between quantities and various constants, under non-restrictive modelling assumptions, enabling the representation of most possible interventions.  

We perform an analysis of variation of reproduction numbers and generation times. While the variation of the former is a main focus of numerous studies, the variation of the latter is often neglected. 
In particular, we highlight that variations of the generation time distribution, if neglected, cause bias in the estimate of reproduction numbers, in some scenarios the bias is significant, while in others it is negligible.  When doing inference, it is thus crucial to investigate the extent of variations of the generation time distribution  in order to determine whether an updated estimate is necessary to avoid significant biases.

 While the general formula \eqref{eq:gen_time} is a valuable theoretical tool for this analysis, we also present  realistic, although  not fitted to real data, examples in order to illustrate the possible variations and draw attention to the problem. 
 We  show that in some cases variations of generation times cause significant biases in the estimation of reproduction numbers, which are crucial for example to establish control measures.
 Because of the general nature of our model, the tools we present can be used in numerous scenarios, combined with real data, to investigate the extent of the variation of the generation time distribution.
 In this way, proper measures can be adopted to avoid biases when needed, that is, either an estimate of the actual generation time distribution is obtained,  or variations of generation time distribution are taken into account in the model itself. 

Several challenges remain to be solved and interesting new directions are open for future work.
In particular,  including a more precise model for contact tracing would sharpen the approximation while requiring additional structure to be included in the model.
Furthermore, an in-depth study of how individual variation affects the generation time remains to be done, 
for example how  the variation in $C_1$, $X_1$ and $X_2$ affects the efficacy of different preventive measures. One such feature of relevance would be to analyse which preventive measures are most efficient in detecting superspreaders, which could be modeled by having large variance of $C_1$, and in particular if effective contact tracing is the best method for reducing their effect on an outbreak.

Finally, the random infectivity profiles in this paper are suitable to represent the potentially different infectiousness and severity of symptoms related to different variants of SARS-CoV-2, thus  enabling an analysis of the impact of new variants on reproduction numbers and generation times, while taking into account the effect of preventive measures. A theoretical analysis, which has been tackled in some specific cases, e.g. \cite{hart2021c}, would be relevant for the planning of control measures in connection with the emergence of new variants of interest and is left for future work. 

\section*{Acknowledgements}
The authors thank the anonymous referees for their constructive comments.
T.B. and M.F. are grateful to the Swedish Research Council (grant 2020-04744) for financial support. G.S.T. acknowledges the MIUR Excellence Department Project awarded to the Department of Mathematics, University of Rome Tor Vergata (CUP E83C18000100006).

\section*{Supplementary material}
\renewcommand{\thesection}{S}
 
 \subsection{Disease spreading before and after symptoms }
According to the definition of random time of symptoms onset, which is given in Section \ref{sect:random_sympt}, it is possible to provide  explicit formulas for $R^{(1)}$ and $R^{(2)}$,  the average number of individuals infected by a typical infectious individual before and after symptoms onset, respectively.
By definition,

 \begin{align*}
     R^{(1)}&=
\E{\int_0^\infty
C_1 X(t) G_{T_S}(t)
dt},
 \\   
     R^{(2)}&=
    \E{ \int_0^\infty
C_2 X(t) 
\left(1-G_{T_S}(t)\right)
dt},\end{align*}
where
    $$
    G_{T_S}(t)= \Prob{T_S >t}=e^{-\int_0^t a_S  X(u) du}.
    $$
Since
$\int_0^\infty X(t)G_{T_S}(t)dt
=-\frac{1}{ a_S }
[e^{- a_S \int_0^\infty X(u)du}-1]$,
we obtain
     \begin{align*}
     R^{(1)}&=
    \E{ C_1
    \frac{1}{ a_S }
    [1-e^{- a_S \int_0^\infty X(u)du}]} ,  
     \\   
     R^{(2)}&=
    \E{
    C_2 \left\{ 
    \int_0^\infty
    X(t) dt + \frac{1}{ a_S }
    [e^{- a_S \int_0^\infty X(u)du}-1]
    \right\}
    }.  
    \end{align*}
    
\subsection{Vaccination}
 
In order to easily illustrate how the infectivity process changes due to vaccination, let us assume, for now,  that 
 the vaccine response is deterministic, i.e. for each individual $A=a\in[0,1]$ and $B=b\in[0,1]$.  
First of all, 
the infectivity process is reduced by a factor $av+1-v$ because of the reduced susceptibility introduced in the population by the vaccine. 
Furthermore, the probability that a randomly chosen  infectious individual is vaccinated is $\frac{av}{av+1-v}$,  in this case  the infectivity profile is further reduced by a factor $b$.  
Denote by $V$ a Bernoulli random variable that is equal to $1$ if the infectious individual is vaccinated, i.e. with probability $\frac{av}{av+1-v}$.
When vaccination is in place, the  infectivity process is thus 
    \begin{equation*}
    \lambda_E(t)=
    [ Vb +1- V]\lambda(t)  [av+1-v ]=
    [ 1-V(1-b)]\lambda(t)  [1-v(1-a) ],
    \end{equation*}
yielding the infectivity function
    $
    \beta_E(t)=
    \E{\rho_V}\beta(t)$,  with 
    $
    \rho_V= [ 1-V(1-b)] [1-v(1-a) ] 
    $ 
and
    $
    \E{\rho_V}=abv +1-v
    $.

If we remove the assumption of a deterministic response, similar formulas hold, involving the expected values of the response random variables, but still leaving the generation time distribution unchanged.

We conclude this subsection by discussing the limitations of our initial assumption that the vaccine may reduce the infectiousness profile of a vaccinated infectious individual by a multiplicative factor, without changing its shape.
In fact,  vaccinated infectious individuals might have an infectious period of a different length, compared to unvaccinated infectious individuals,  also, they might be less likely to show symptoms and might have a different behaviour, for example due to the decreased severity of the disease or to the security provided by the vaccine.  Therefore, in reality,  their infectivity profile might be  different, not only by a multiplicative factor, from the unvaccinated infectivity profile.
 A differently shaped infectivity profile of vaccinated individuals could be included to make the vaccination model more realistic, by incorporating ad hoc assumptions on the infectivity profile, based on  available evidence on a specific disease.
However,
we point out that a differently shaped infectivity profile of vaccinated individuals would  contribute to the formula for $\beta_E$ with weight $\frac{abv}{abv+1-v}$, which ideally is a small quantity if the vaccine is effective.  
This heuristically hints that, under more general assumptions on the infectiousness profile of vaccinated individuals,  variations of the generation time distribution due to vaccination are anyway expected to be small for an effective vaccine.

\subsection{Contact tracing}
As mentioned in Section \ref{sect:model}, the epidemic model in this paper corresponds, under certain conditions, to a branching process. Contact tracing could be simulated exactly by superimposing a contact tracing process on the branching process. That is, when an individual is detected, the infectivity profile of the consequently  traced individuals is reduced and the original branching tree might be pruned.  
Whereas, our approximation  allows to  directly simulate a branching process with discounted birth rates, as shown in the following, incorporating the effect of contact tracing, instead of considering the original branching process with superimposed contact tracing.
In particular, we define an infectivity process that accounts for the possible reduction of contact activity due to the individual being contact traced by defining the conditional contact tracing rate, $\alpha_{CT}$, which increases the detection rate and thus discounts the birth rate in the corresponding branching process. 
This section contains a rigorous derivation of the  contact tracing rate and further discussion on the assumptions.

Consider an infectious individual, called infector from now on,  and the process $N=\{N(t)\}_{t\geq0}$ counting the number of infections caused by them. 
That is, $N(t)$ is the number of individuals the infector has infected  by time $t$. Time is counted since the infection of the infector.  
For $i=1,\dots, N(t),$ let
$T^{(i)} $ be the time between the infection of infectee number $i$ and the infection of the infector, and let $T_D^{(i)}$ be the time between the infection and detection of infectee number $i$.
It is known that, given $N(t)$ and $\lambda$, the $T^{(i)}$'s are independent and  identically distributed random variables with probability density function  $\frac{\lambda(t)}{\int_0^t\lambda(u) du}\mathbb{I}_{[0,t]}$, see e.g. \cite{snyder1991}. 
We now use a 
recursive argument, that is, assuming that $T_D$ is already defined and letting 
$\mathcal{L}_t$ be the sigma algebra generated by the process $\lambda$ up to time $t$, we compute the through-infectees contact tracing rate $\alpha_{CT1}$.
To this aim, we compute, conditionally on $\mathcal{L}_t$, 
the probability that the infector has not yet been  contact traced through their infectees by time $t$, i.e. the probability that the time, $T_{CT1}$, at which the infector is contact traced through their infectees, is larger than $t$. This  is  equal to the probability that none of the infectees  has been detected by time $t$,  assuming for now that there is no delay between detection and contact tracing and that it is possible to trace all contacts. 
Letting 
$\bar{G}_{T_D}(t)=\Prob{T_D>t}=\E{G_{T_D}(t)}$ be the unconditional distribution of $T_D$, with corresponding probability density function $\bar{f}_{T_D}=-\frac{d}{dt}\E{G_{T_D}(t)}$, and using the properties of the $T^{(i)}$'s yields
 \begin{align*}
    \Prob{T_{CT1}>t \mid \mathcal{L}_t}&=
    \Prob{\bigcap_{i=1}^{N(t)}\{T_D^{(i)}>t-T^{(i)}\} \mid \mathcal{L}_t}\\
    &=
    \E{
    \Prob{\bigcap_{i=1}^{N(t)}\{T_D^{(i)}>t-T^{(i)}\}
    \mid N(t),\{T_i\}_{i=1}^{N(t)}}
    \mid \mathcal{L}_t}
    \\&=
    \E{
    \prod_{i=1}^{N(t)}
    \E{\bar{G}_{T_D}(t-T^{(i)})\mid \lambda,N(t)}
    \mid \mathcal{L}_t}
     \\&=
     \E{
    \E{\bar{G}_{T_D}(t-T^{(1)})\mid N(t)}^{N(t)}
    \mid \mathcal{L}_t}
    \\&=
     \E{
    \left(
    \int_0^t \bar{G}_{T_D}(t-v) 
    \frac{\lambda(v)}{\int_0^t\lambda(u) du}
    \right)^{N(t)}
    \mid \mathcal{L}_t}
    \\&=
    \exp\left(\left(\int_0^t \bar{G}_{T_D}(t-v) 
    \frac{\lambda(v)}{\int_0^t\lambda(u) du}dv-1\right)\int_0^t\lambda(u)du \right)
    \\&=
    \exp\left(-\int_0^t\lambda(v)\left(1-\bar{G}_{T_D}(t-v) 
    \right)dv \right)
    \end{align*}
 where we have also used that $N(t)$ given $\mathcal{L}_t$  is Poisson$(\int_0^t\lambda(u)du)$, with  conditional probability generating function  $\E{s^{N(t)}\mid\mathcal{L}_t}=\exp\left((s-1)\int_0^t\lambda(u)du\right)$.
Therefore, 
    $$
    \int_0^t \alpha_{CT1}(v)dv=
    \int_0^t\lambda(v)\left(1-\bar{G}_{T_D}(t-v) 
    \right)dv, 
    $$
and differentiating yields
    $$
     \alpha_{CT1}(t)=
    \int_0^t\lambda(v)\bar{f}_{T_D}(t-v) 
    dv .
    $$
Let $p$ be the contact tracing success probability, that is, the probability that a contact can be actually traced.
Since we are  interested in the contact tracing rate only until detection occurs,   in which case $\lambda(t)=C_1 X(t)$, 
we obtain, conditionally on $C_1$ and $\mathcal{X}_t$,
\begin{align*}
    \alpha_{CT1}(t) =p \int_0^t C_1 X(v)\bar{f}_{T_D}(t-v) dv 
    = p
    \int_0^t C_1 X(v) dv 
    +p
    \int_0^t C_1 \bar{G}_{T_D}(v)X(t-v)dv,
\end{align*}
using integration by parts for the last equality.
Therefore,
since
$\alpha_D(t)=\alpha_S(t)+\alpha_{scre}(t)+\alpha_{CT1}(t)+\alpha_{CT2}(t)$,

    \begin{align}
    \label{eq:G_tau}
    G&_{T_D}(t)\\
    \nonumber
    &=\exp\left(
    -\int_0^t [\alpha_S(u)+\alpha_{scre}(u) +\alpha_{CT2}(u)] du 
    - p C_1 \left[  \int_0^t X(u) du + \int_0^t  \bar{G}_{T_D}(u)X(t-u)du \right]
    \right)
    \end{align}
In order to compute $G_{T_D}$,  the conditional distribution of $T_D$, by using the equation above,  $\bar{G}_{T_D}$ should be computed first.  
Assuming that an explicit expression for $ \alpha_{CT2}$ is known and taking the expectation of the equation above  yields an integral equation for $\bar{G}_{T_D}(t)$ which should be solved numerically, then, by plugging the numerical solution in   \eqref{eq:G_tau},  $ {G}_{T_D}$ can be computed.

 Instead, to obtain an explicit expression for $G_{T_D}$, by making an approximation, it is possible to  assume that contact tracing  stops after one step, which corresponds to assuming that $\bar{G}_{T_D}(t)$ in \eqref{eq:G_tau} is replaced by 
    $$
     \E{\exp\left(-\int_0^t[\alpha_S(u)+\alpha_{scre}(u)]du\right)} .
    $$
In this case, \eqref{eq:G_tau} provides an explicit expression for the conditional distribution of the detection time, and  the integral equation is avoided.  

We can also  consider a delay between the moment an infectee is detected and their infector is contact traced.
Letting $d$ be the probability density function of the delay time, and $\bar{f}_{T_D}^{d}(t)=\int \bar{f}_{T_D}(t-v)d(v)dv$, the through-infectees contact tracing rate becomes
    \begin{align*}
    \alpha_{CT1}(t) = p \int_0^t C_1 X(v)\bar{f}^{d}_{T_D}(t-v) dv .
    \end{align*}
 If the delay, $d\in \mathbb{R}_{>0}$, is deterministic, we obtain 
  \begin{align*}
    \alpha_{CT1}(t) = p \int_0^{t-d} C_1 X(v)\bar{f}_{T_D}(t-d-v) dv .
    \end{align*}

We conclude this section with a discussion on the through-infector contact tracing rate. 
In fact, while we have assumed it constant, this rate  varies with time, depending on how long has passed since the infector was infected.  Thus, in a rigorous derivation of this rate, the generation time distribution should be involved. At the same time, the expression of this rate is needed to express the generation time distribution, leading to a loop of definitions that is not easy to disentangle in the current framework. 
In \cite{scarabel2021} a deterministic model is used, and an integral equation for the through-infector contact tracing rate  is derived, however, a similar argument does not apply here, because of the additional challenges caused by the random infectivity profile. 
One of the challenges would consist in describing the infectivity profile of an infector, in fact, such an individual  ought to be more infectious that an average individual, since at  least one infection has occurred.
To address these problems, additional structure in the model would be required.
Instead,  since the rate at which an individual is contact traced through their infector does not depend on their infectivity process, 
we approximate the through-infector contact tracing rate by a constant. The value of the constant depends on the specific setting and disease of interest, see Section \ref{sect:ill_CT} for an illustration.

\subsection{Example: SIR with interventions}
Within the framework described in Example \ref{example:SIRinterventions}, we compute the generation time distribution explicitly as follows.
A first simplification yields
    \begin{align*}
    &\beta_E(t)=
    c \xi \E{ \mathbb{I}_{[0,I]}(t) G_{T_D}(t)},
    \\
    &G_{T_D}(t)=
     \exp\left(
      - (a_S \xi + \sigma+ p a_{CT2}) (t \wedge I ) 
     - p c \xi \int_0^{t-d}   \mathbb{I}_{[0,I]}(u)[1-y (t-d-u) ]du   
     \right), \\
     &y(t)= \E{\exp\left(-(a_S \xi + \sigma) (t \wedge I ) \right)}
     .
    \end{align*}
Using that $I\sim \text{Exp} (a_I)$, yields
    $$
    1- y(t)= \frac{a_S \xi + \sigma }{a_I +a_S \xi + \sigma } \left[ 1- e^{ -(a_I +a_S \xi + \sigma)t}\right].
    $$
Therefore,
    \begin{align*}
    G_{T_D}(t)= &
    \exp\left(
      - (a_S \xi + \sigma+ p a_{CT2}) (t \wedge I ) 
       \right)
       \exp\left(
      - p c \xi \frac{a_S \xi + \sigma }{ a_I +a_S \xi + \sigma}((t-d) \wedge I ) 
       \right)
       \\
       & \cdot \exp\left(
      - p c \xi \frac{a_S \xi + \sigma }{ (a_I +a_S \xi + \sigma)^2}
      \left[ 
      e^{ -(a_I +a_S \xi + \sigma)(t-d)} - e^{ -(a_I +a_S \xi + \sigma)[t-d-I]^+}
      \right]
       \right),
       \end{align*}
and thus
        \begin{align*}
       \beta_E(t)= &
       c \xi \exp\left(
      - (a_S \xi + \sigma+ p a_{CT2}) t 
       \right)
       \exp\left(
      - p c \xi \frac{a_S \xi + \sigma }{ a_I +a_S \xi + \sigma}(t-d) 
       \right)
       \\
       & \cdot \exp\left(
      - p c \xi \frac{a_S \xi + \sigma }{ (a_I +a_S \xi + \sigma)^2}
      \left[ 
      e^{ -(a_I +a_S \xi + \sigma)(t-d)} - 1
      \right]
       \right)
       \Prob{I>t-d}.
     \end{align*}
In order to identify the generation time distribution, we can omit multiplicative constants in the expression above
and obtain, using $ \propto$ to indicate proportionality with respect to $t$,
 \begin{align*}
       \beta_E(t)\propto & \ 
        \exp\left(
      - \left(a_I +  a_S \xi + \sigma+ p a_{CT2} + p c \xi \frac{a_S \xi + \sigma }{ a_I +a_S \xi + \sigma} \right)t
       \right)
       \\
       & \cdot \exp\left(
      - p c \xi \frac{a_S \xi + \sigma }{ (a_I +a_S \xi + \sigma)^2}
      e^{ (a_I +a_S \xi + \sigma)d}
      e^{ -(a_I +a_S \xi + \sigma)t}
        \right).
     \end{align*}

We recall that the Exponential-Gamma distribution, see e.g. \cite{pinheiro2015}, is a generalisation of the Gumbel distribution with probability density function 
    $$
    f_{EGa}(t;\mu_{EG},\sigma_{EG},\alpha_{EG})=
    \frac{1}{\Gamma(\alpha_{EGa})}
    \frac{1}{\sigma_{EG}} 
    \exp\left(
    - \alpha_{EGa} \frac{t- \mu_{EG} }{\sigma_{EG}}
    \right)
    \exp\left(
    - 
    e^{-\frac{t- \mu_{EG} }{\sigma_{EG}}}
        \right).
    $$
The Gumbel distribution is recovered when $ \alpha_{EG}=1$.
Therefore, by matching the expressions above, it follows that  the generation time distribution in this example is a truncated version (with support on the positive real numbers) of an Exponential-Gamma distribution, with parameters
    \begin{align*}
      \mu_{EG}&=\frac{1}{a_I +a_S \xi + \sigma}
      \log\left(  \frac{p c \xi(a_S \xi + \sigma) }{ (a_I +a_S \xi + \sigma)^2}\right)
      + (a_I +a_S \xi + \sigma) d,
      \\
      \sigma_{EG}&= \frac{1 }{ a_I +a_S \xi + \sigma},
      \\
      \alpha_{EG}&= 1+ \frac{p a_{CT2} }{ a_I +a_S \xi + \sigma}
      + \frac{p c \xi(a_S \xi + \sigma) }{ (a_I +a_S \xi + \sigma)^2}.
    \end{align*}

\bibliographystyle{plain}

\end{document}